# Broadband voltage rectifier induced by linear bias dependence in CoFeB/MgO magnetic tunnel junctions

*M. Tarequzzaman*[1,2], *A. S. Jenkins*[1], *T. Böhnert*[1], *J. Borme*[1], *L. Martins*[1], *E. Paz*[1], *R. Ferreira*[1] *and P. P. Freitas*[1,2]

[1]INL - International Iberian Nanotechnology Laboratory, Ave. Mestre Jose Veiga, 4715-330, Braga, Portugal

[2]Physics Department, Instituto Superior Tecnico (IST) – Universidade de Lisboa, 1000-029, Lisbon, Portugal

## ABSTRACT

In this paper the perpendicular magnetic anisotropy (PMA) is tailored by changing the thickness of the free layer with the objective of producing MTJ nano-pillars with smooth linear resistance dependence with both in-plane magnetic field and DC bias. We furthermore demonstrate how this linear bias dependence can be used to create a zero-threshold broadband voltage rectifier, a feature which is important for rectification in wireless charging and energy harvesting applications. By carefully balancing the amount of PMA acting in the free layer the measured RF to DC voltage conversion efficiency can be made as large as 11%.

## INTRODUCTION

The spin-transfer torque (STT) commonly reported in magnetic tunnel junctions (MTJ) or spin-valves occurs when a spin-polarized current[1–5] is applied and has the ability to reverse the magnetic orientation and induce persistent magnetization oscillations, therefore allowing one to manipulate magnetization of a nanoscale magnetic object[6,7]. This can lead to various applications such as magnetoresistive random-access memory (MRAM)[8,9], magnetic sensors[10,11] and ultra-tunable microwave oscillators[12,13]. Later it was shown that, if the DC bias is replaced with a radiofrequency (RF) current through the MTJ nanopillar, a rectified DC voltage can be generated by the so-called spin-torque diode effect, where the STT associated with the applied RF current results in the magnetisation of the free layer of the MTJ nanopillar oscillating as a function of time. The resultant oscillating resistance is mixed with the applied RF current and generates a DC rectified voltage ($V_{rect}$) across the MTJ[14–16]. One possible application of this effect is the spin-torque microwave detectors (STMDs)[6,15,17]. A recent study suggests that high RF detection sensitivities in MTJ based nanopillars can be higher than semiconductor Schottky diode detectors by using a free layer with perpendicular magnetic anisotropy (PMA)[17–19].

In this paper we show how the magnetisation reversal of the free layer, and therefore the resistance of the magnetic tunnel junction, can be tailored via the introduction of a perpendicular magnetic anisotropy. This tailoring can be used to create a system which displays a linear resistance variation as a function of the magnetic field or, more critically, the DC bias current. This linear dependence on the bias is used to demonstrate a zero-threshold broadband voltage rectifier, which unlike conventional diodes, remains operational even with very weak RF excitation currents, as those typically generated by many energy harvesting devices. The fact that all the results were obtained at very low magnetic fields (which can be potentially zero for future devices) and all the measurements were carried out at room temperature, makes this result well suited for real-world applications.

# EXPERIMENTAL METHOD

In order to quantify the PMA in Ta/CoFeB/MgO system a set of six samples were deposited on a (10×5 mm$^2$) Si/100 nm Al$_2$O$_3$ wafer. These samples incorporated a single CoFeB layer sandwiched in the stack, Ta/CoFeB/MgO layers (50 Ta/ **1.0-2.0** Co$_{0.4}$Fe$_{0.4}$B$_{0.2}$/MgO [3.0 Ω·µm$^2$]/ 5 Ta/ 7 Ru) (thickness in nanometers), which was deposited with a variable thickness of CoFeB between 1.0 nm to 2.0 nm. After the deposition, the samples were annealed at 330°C, for 2 hours, under a 1 T magnetic field oriented along the easy axis direction, defined by a magnetic field applied during the deposition.

These samples were then characterized in a VSM, extracting the in-plane (IP) and out-of-plane (OOP) magnetization versus applied field (*M-H* curve). From these measurements, the thickness ($t_{nominal}^{CoFeB}$) dependence of the saturation magnetization ($m_s$) indicates an approximately 0.52±008 nm thick magnetically dead layer in CoFeB inserted the Ta/CoFeB/MgO tri-layer as shown in the Fig. 1(a). The effective CoFeB thickness ($t_{eff}^{CoFeB}$) is defined by the dead layer thickness subtracted from the nominal thickness and plotted against the total effective anisotropy energy ($K_{eff}$) as shown in Fig. 1(b). The total effective anisotropy energy ($K_{eff}$) is expressed by[20],

$$K_{eff} = K_b - \frac{M_s^2}{2\mu_0} + K_i / t_{eff}^{CoFeB} \qquad (Eq.\ 1)$$

Here $K_b$ is the bulk magnetocrystalline anisotropy and $K_i$ is the interfacial anisotropy. From the data fitting via Eq. 1, ($t_{eff}^{CoFeB}$ dependence of $K_{eff} \times t_{eff}^{CoFeB}$) plotted in Fig. 1(b), different anisotropy contribution can be evaluated. Here ($K_b$-$M_s^2/2\mu_0$) corresponds to slope of linear fit and $K_i$ is the vertical intercept of the linear fit. The bulk contribution is consistent with the demagnetization (-$M_s^2/2\mu_0$) indicating that $K_b$ is negligible and $K_i$ is determined to be 0.60±0.07 mJ/m$^2$. Thus, the PMA in Ta/CoFeB/MgO system is entirely due to the interfacial anisotropy[21,22]. This value of $K_i$ is consistent with the high annealing temperature (330$^0$C), which is known to reduce the PMA[21,23] of Ta/CoFeB/MgO system.

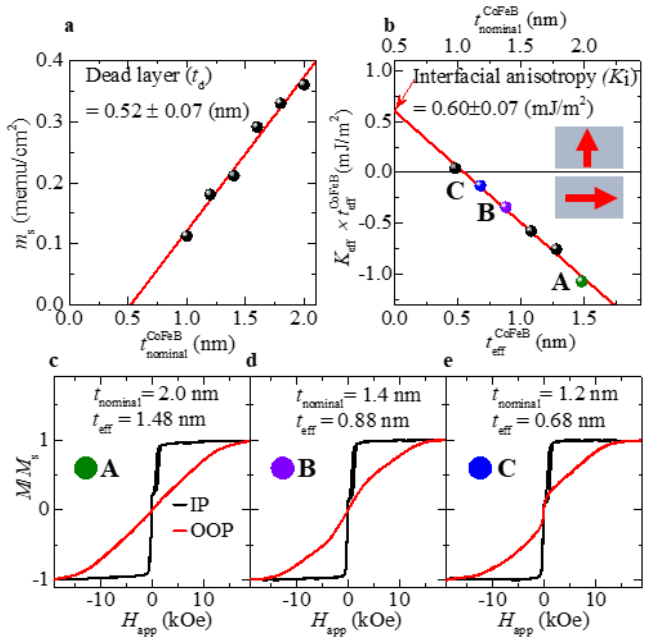

Fig. 1(a) Saturation magnetization as a function of nominal CoFeB thickness ($t_{nominal}^{CoFeB}$). The linear data fitting indicates 0.52±0.07 nm magnetic dead layer. **(b)** Dependence of the $K_{eff}$ vs $t_{eff}^{CoFeB}$ product on the effective thickness ($t_{eff}^{CoFeB}$) for the films. From linear extrapolation of the data fitting, intercept in the vertical axis corresponds to $K_i$ and slopes represents ($K_b$-$M_s^2/2\mu_0$). **(c-e)** In-plane (black lines) and out-of-plane (red lines) magnetization curves of unpatterned samples (10×5 mm$^2$) with FL thicknesses A, B and C. All samples were annealed at 330°C, 2 hours, and 1 T magnetic flux along the easy axis and measured in VSM.

Three MTJ stacks (50 Ta/ **X** Co$_{0.4}$Fe$_{0.4}$B$_{0.2}$/MgO [3.0 Ω·µm$^2$]/2.2 Co$_{0.4}$Fe$_{0.4}$B$_{0.2}$/0.85 Ru/2.0 Co$_{0.7}$Fe$_{0.3}$/20 Ir$_{0.2}$Mn$_{0.8}$/ 15 Ru) (thickness in nanometers) were then deposited on a 200 mm Si/100 nm Al$_2$O$_3$ wafer. As before, the wafers were annealed at 330°C, for 2 hours, under a 1 T magnetic field oriented along the easy axis. The nominal thicknesses of the free Co$_{0.4}$Fe$_{0.4}$B$_{0.2}$ layer (**X**) were 2.0 nm, 1.4 nm and 1.2 nm. Subtracting from the nominal free layer (FL) thicknesses the dead layer measured in the previous set, these three samples incorporate a free layer with effective magnetic thicknesses of $t_{eff}$ =1.48 nm, $t_{eff}$ = 0.88 nm and $t_{eff}$ = 0.68 nm, respectively. VSM measurements performed in bulk coupon samples extracted from the wafers (10×5 mm$^2$) are shown in Fig. 1 (c-e), displaying measurements of the IP (black lines) and OOP (red lines) magnetization curves. The out-of-plane (OOP) measurements clearly show that the perpendicular saturation magnetic field ($H_K$) increases by decreasing the FL thickness, consistent with previous single layer CoFeB measurements and reported publications[20]. The FL thickness **A**, exhibits a strong in-plane (IP) uniaxial anisotropy as expected for thick $Co_{0.4}Fe_{0.4}B_{0.2}$ layer. Whereas, for FL thickness **B**, the easy axis can be seen to tilt slightly towards the OOP direction resulting in an increase of $H_K$. When the FL thickness is further reduced to **C**, the FL tilts further towards the OOP magnetization direction. This is due to the enhancement of PMA at the interface between CoFeB/MgO, as the thickness of CoFeB layer is reduced[20,24].

The samples (200 mm wafers) were then patterned into elliptical nanopillar shapes (with axial diameters of 75 nm × 110 nm) by electron beam lithography combined with ion milling system. After the device fabrication, the devices were contacted in a 4-point contact lead geometry.

In Fig. 2, the in-plane magnetic field dependence of the nano-pillar resistance is measured under a low bias ($I_{bias}$ = 0.01 mA) for the three FL thicknesses. The transfer curves measured for all the FL thicknesses are slightly displaced from zero in-plane magnetic field due to the internal fields acting on the free layer: a strong ferromagnetic coupling[4] between the reference layer and FL and the stray field of the patterned reference SAF acting on the free layer. For Fig. 2 (a), the resistance can be seen to exhibit sharp switching as a function of the magnetic field with a pronounced hysteretic behaviour which is typical of systems with a strong in-plane easy axis[25].

Moving to Fig. 2 (b), as the FL thickness is reduced the switching remains relatively sharp but it is no longer hysteretic. The system clearly shows an in-plane easy axis but the addition of the PMA results in a small tilting of the magnetisation out of plane as the effective magnetic field approaches zero[26] (i.e. when the applied field compensates internal fields acting on the free layer, which happens when $H_{app}$ = 68 Oe). This tilting is relatively small but is evidently enough to allow a smoother transition between the parallel and anti-parallel states and, critically, removes the hysteresis from the switching.

For the thinnest free layer (**C**), shown in Fig. 2(c), switching is more gradual than for the other thicknesses, as the FL is tilted further towards the out-of-plane direction when the in-plane field arrives at $H_{eff}$ = 0 mT[26], and a strong magnetic field is now needed to saturate the FL magnetization along the in-plane direction.

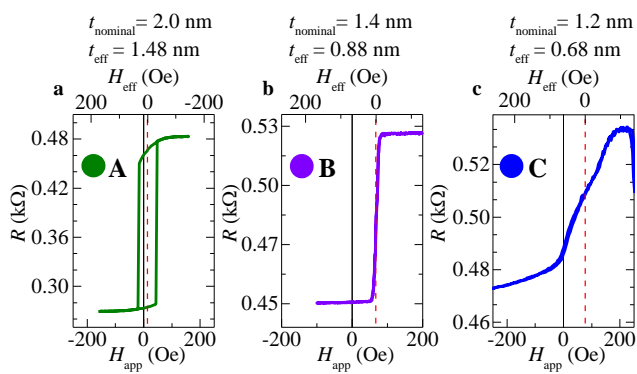

Fig. 2 (a-c) Electrical resistance as a function of the in-plane magnetic field with an applied bias current of 0.01 mA for the three FL thicknesses (**A, B and C**). Red dash line represents $H_{eff}$ = 0.

Subsequent to the discussion of in-plane magnetic fields at a low DC current, the DC bias current dependence is shown more thoroughly in Fig. 3 (a-c), where a certain magnetic field ($H_{app}$) is applied along the in-plane easy axis in order to measure the switching in $H_{eff}$ = 0 (Red dash line in the Fig. 2) and the resistance is plotted as a function of the applied DC current. For Fig. 3 (a), where the PMA is negligible (i.e **A**), the switching with the DC current can be seen to be similar to that by an in-plane magnetic field, i.e. sharp and hysteretic, as the STT originating from the spin polarised current is switching the FL along the easy axis[27].

However, for Fig. 3(b), where a weak PMA leads to a very slight tilting out-of-plane (i.e. **B**), the resistance bias dependence can be seen to be linear, echoing the same linearity that was observed for the magnetic field dependence. The system exhibits a non-hysteretic linear bias dependence which is centered at $I_{bias}$ = 0 mA, with a gradient of $dR/dI$ = 397 Ω/mA. Non-hysteretic switching can also be seen when the system exhibits a relatively strong PMA (i.e. **C**) as shown in Fig. 3 (c), but the strongly tilted out-of-plane system is less linear as a function of the bias current, as well as having a reduced $dR/dI$.

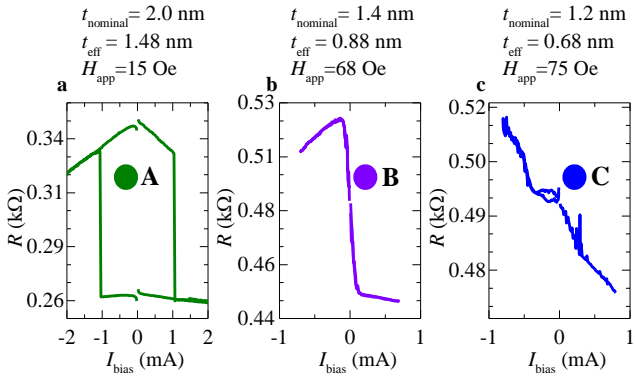

Fig. 3 **(a-c)** Current-induced magnetization switching in the free layer thicknesses (**A, B and C**). The measurement were carried out in $H_{eff}$ = 0 applying a magnetic field **(a)** $H_{app}$= 15 Oe **(b)** $H_{app}$= 68 Oe & **(c)** $H_{app}$= 75 Oe along the in-plane easy axis.

The non-hysteretic linear behaviour of the resistance centred at zero bias, which is presented so clearly in the system with the weak PMA (i.e. **B**) in Fig. 3 (b), makes these magnetic tunnel junctions ideal candidates for potential broadband voltage rectifiers. The incoming high frequency current will result in a variation in the resistance, due to the linear $dR/dI$. The mixing of the high frequency current passing through the nano-pillars and the resultant high frequency resistance oscillation will result in a voltage, the root mean square of which will be non-zero. As long as this current variation is centred at $I_{bias}$ = 0, as is the in Fig. 3 (b), there will be zero threshold in order to observe the rectified voltage.

To illustrate this effect, which can broadly be described as a modulation-based spin torque diode effect[16], an RF current was generated by a high frequency source and passed through a bias tee to the sample, and the DC component of the voltage was measured using a voltmeter. The output can be seen in Fig. 4 (a) as a function of the frequency of the RF current for three thickness (**A, B and C**). By comparing the measured result for three different thicknesses, it is clear that the largest rectified voltage is observed for the weak PMA with free layer thickness **B** (around 37 mV when an RF current of around 0.65 mA is applied through the MTJ). This is due to the sharpness of the transition in the linear $dR/dI$ (as shown in Fig 3 (b)). A smaller, but still considerable rectified voltage can still be observed for case with a strong PMA, i.e. **C**, which is reduced due to the reduced gradient observed in Fig 3 (c) (13 mV for same RF current). A very weak effect can be seen in the system with negligible PMA, i.e. **A**, but is orders of magnitude smaller than for the systems with a PMA (weak and strong PMA).

This modulation effect can be thought of as a non-resonant spin torque diode effect. The rectified voltage depends on the effect of the modulation source frequency ($f_{source}$) and the relaxation frequency ($f_p$) of the system. The relaxation frequency ($f_p$) is defined as $f_p = \Gamma_P/\pi$[28], where the amplitude relaxation rate ($\Gamma_P$) is an intrinsic parameter which determines the timescale over which the system can respond to an external stimulus. The sign of the resultant voltage depends on the sign of the linear behaviour observed in Fig. 3 (b and c) (i.e. negative slope results in negative voltage). Similar behaviour has been found in a variety of shapes and sizes.

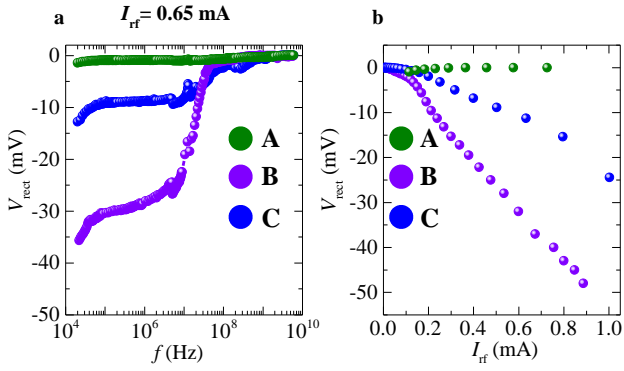

Fig. 4 (a) Measured voltage $V_{rect}$ as a function of the applied RF current of $I_{RF}$=0.65 mA at $H_{eff}$ = 0 for three FL thicknesses (**A**, **B** and **C**). The $V_{rect}$ were measured across an MTJ at zero applied DC bias. The cut-off frequency of the system is around $10^7$ Hz, from where the system starts relaxing rapidly. **(b)** Measured rectified output voltage as a function of the applied RF current.

The rectified voltage can be seen to clearly depend on the incoming RF source frequency. For low frequencies ($10^5$ Hz to $10^7$ Hz), (where $f_{source} < f_p$), the system runs in a free stationary trajectory and the subsequent mixing results in the generation of a large output voltage ($V_{rect}$). For larger frequencies (>$10^7$ Hz), the source frequency ($f_{source}$) starts to become higher than the relaxation frequency ($f_p$) and the ability for the modulation current to generate sustained oscillations is strongly reduced, generating a very low rectified voltage. This roll-off in the modulation when $f_{source} > f_p$ has been previously reported with time domain measurements in ref. [28]. The additional features (i.e. small peaks and troughs) observed in the voltage are due to the measurement chain and the frequency source and are not believed to relate to the MTJ itself. The fact that, this broadband rectification effect shows a large output voltage generation over several decades of frequency in magnetic tunnel junctions, with zero DC bias and small, potentially zero, applied magnetic fields, making them exciting candidates as potential wireless chargers.

To quantify the rectification effect measured for the three FL thicknesses of **A**, **B** and **C**, the measured rectified output voltage is plotted as a function of the applied RF current as shown in Fig. 4 (b). Among the three FL thicknesses, the weak PMA shows highest $V_{rect}$ of around 48 mV for an applied RF current amplitude $I_{RF}$ = 0.89 mA. The RF to DC voltage conversion efficiency ($\eta$) of a device with a resistance $R$ can be express as:

$$\eta = V_{rect}/(I_{RF} \times R) \quad \text{(Eq. 1)}$$

From the Eq. 1, the RF to DC conversion efficiency for weak PMA (thickness **B**) is found to be 11%. Note that for both systems with PMA (i.e. **B** and **C**), the dependence of the rectified voltage on the applied current is roughly linear, which originates from the linear behaviour observed in Fig. 3. The observed rectification effect makes these systems attractive candidates for RF-DC rectification applications, due to the passive zero-threshold broadband voltage which is generated by an incoming RF current.

This broadband rectification effect can be well suited for another potential application namely as a power sensor, where the power of the incoming signal can be clearly extracted by measurement of the rectified voltage. A clear advantage of this type of power sensor is that it has a relatively low sensitivity to small frequency changes due to the broadband nature of the response, and can work for low power signals due to the zero-threshold nature of the rectification effect.

# CONCLUSION

In summary, we have demonstrated that the nature of the switching of the free layer in a magnetic tunnel junction can be precisely tailored by the introduction of a PMA. For systems with a weak PMA we have observed a sharp linear variation in the resistance with respect to the applied DC bias, centred at zero DC bias. This linear behaviour with respect the applied current makes these systems exciting candidates as broadband non-resonant zero-threshold voltage rectifiers.

## ACKNOWLEDGEMENTS

The research leading to these results has received funding from the European Union Seventh Framework Programme [FP7-People-2012-ITN] under Grant agreement No. 316657 (SpinIcur) and from Norte2020 project Nanotechnology based functional solutions (NORTE-01-0145-FEDER-000019). M. Tarequzzaman thanks the European Union for funding the European Union Seventh Framework Programme [FP7-People-2012-ITN].